\documentclass[final,3p,times,authoryear]{elsarticle}

\usepackage{amssymb}
\usepackage{lipsum}

\usepackage{orcidlink}
\usepackage{ulem}

\usepackage{lineno}

\journal{New Astronomy}

\begin{document}

\begin{frontmatter}



\title{A fast photometric image alignment algorithm with row and column means}


\author[keylab]{Jie Zheng~\corref{me}~\orcidlink{0000-0001-6637-6973}}
\affiliation[keylab]{organization={CAS Key Laboratory of Optical Astronomy, National Astronomical Observatories, Chinese Academy of Sciences},
            addressline={}, 
            city={Beijing},
            postcode={100101}, 
            state={Beijing},
            country={China}}

\author[lsnu]{Linqiao Jiang~\orcidlink{0000-0002-2760-8492}}
\affiliation[lsnu]{organization={School of Mathematics and Physics, Leshan Narmal University},
            addressline={}, 
            city={Leshan},
            postcode={614000}, 
            state={Sichuan},
            country={China}}

\author[keylab]{Jianfeng Tian~\orcidlink{0000-0003-0672-3579}}

\cortext[me]{jiezheng@nao.cas.cn}

\begin{abstract}

This paper introduces an astronomical image alignment algorithm. This algorithm uses the means of the rows and columns of the original image for alignment, and finds the optimal offset corresponding to the maximum similarity by comparing different offsets between images. The similarity is evaluated by the standard deviation of the quotient divided by the means. This paper also discusses the theoretical feasibility of this algorithm. Through practical testing, it has been confirmed that the algorithm is fast and robust.
\end{abstract}



\begin{keyword}
methods: data analysis \sep techniques: image processing \sep binaries: eclipsing



\end{keyword}

\end{frontmatter}



\section{Introduction}

Astronomy is a science based on observation, and photometry observation is one of the important methods in astronomical research. The reduction of observation images is the most fundamental work. In most time-domain observation data reduction, the absolute celestial coordinates of objects are not important. Especially when processing light-curves from data for specific targets \citep{QLCP}, such as binary stars, exoplanets, and supernovae, the relative displacement between images is crucial. On the other hand, many astronomical observations are currently conducted on equatorial telescopes, with displacement between images but no rotation. The rotator of an altazimuth telescope would help to keep the rotation angle of images. Many tools can perform relative alignment, such as astroalign \citep{2020A&C....3200384B}, a pattern-matching algorithm \citep{1986AJ.....91.1244G}, a point pattern matching program \citep{1992PASP..104..301M}, and so on. But their work is all based on star catalogs extracted from images. Some of these tools can be easily integrated into our own programs, while others need to be used by calling external programs. If the absolute celestial coordinates are required, tools like SCAMP \citep{scamp} and astrometry.net \citep{am.net} will help.

Here a new algorithm is introduced. This algorithm will not call object extraction program, it will relay on the row and column means of the observed images, and even without bias-subtraction and flat-fielding. But it only works on images without field rotation. It can only find the translation between images. And it is very simple, fast, and robust.

This program is published at the gitee: https://gitee.com/drjiezheng/mean\_offset/ .

\section{Algorithms}

\begin{figure}[h]
	\centering 
	\includegraphics[width=0.4\textwidth]{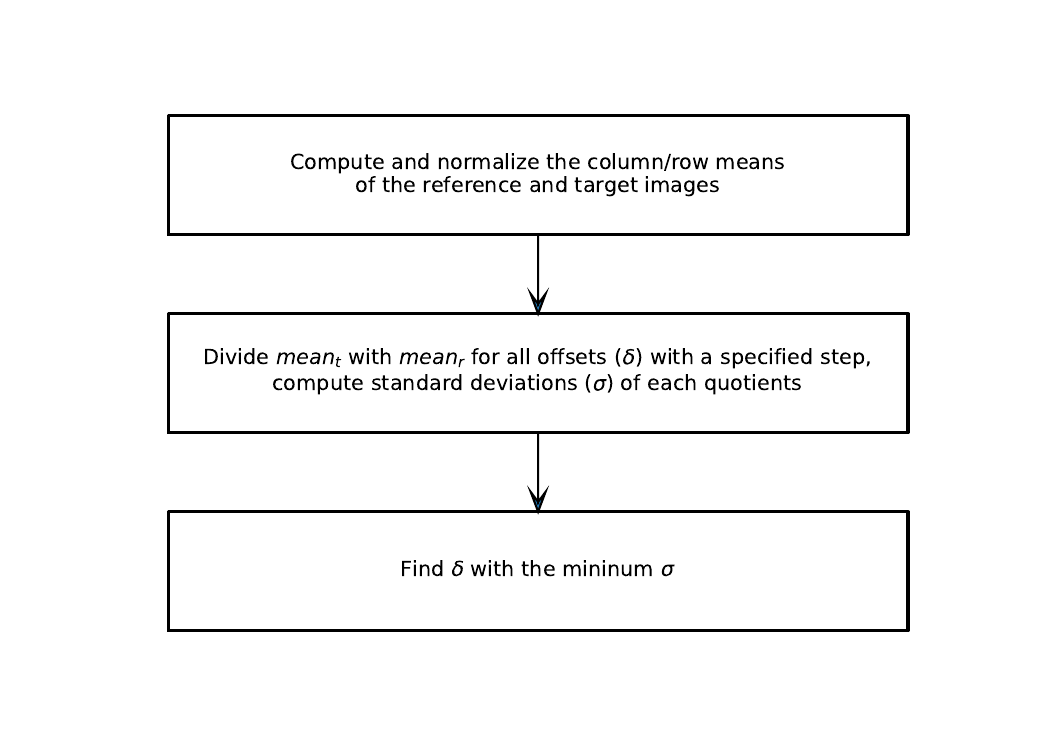}
	\caption{Steps of the algorithm} 
	\label{figStruct}%
\end{figure}

The general structure of the algorithm is shown in Figure \ref{figStruct}. A target image will be aligned with a reference image. At first, the row means ($mean_y$) and column means ($mean_x$) of the reference image and the target image are computed, which are called the $mean_{xr}$, $mean_{yr}$, $mean_{xt}$, $mean_{yt}$.
The similarity between the translated means are measured. Finally, the translated candidate corresponding to the best similarity was chosen as the offset. The x and y offsets are treated independently. 

\subsection{Means of Each Row and Column}


The simple means of the images of each row and column can be computed easily in any programming language.
In this work, the similarity of two sequences was expressed by the standard deviation the quotients of one sequence divided by the other.
The direct average has certain flaws, the exposure time of images will affect the result. Meanwhile, the gradient of the background, including the sky background and the illumination of the telescope, makes the quotients unstable. To remove these side effects, the data were normalized by the sky background. To estimate the sky background, a 3-sigma-clipped parabola fitting was employed.

\begin{figure}[h]
	\centering 
	\includegraphics[width=0.5\textwidth]{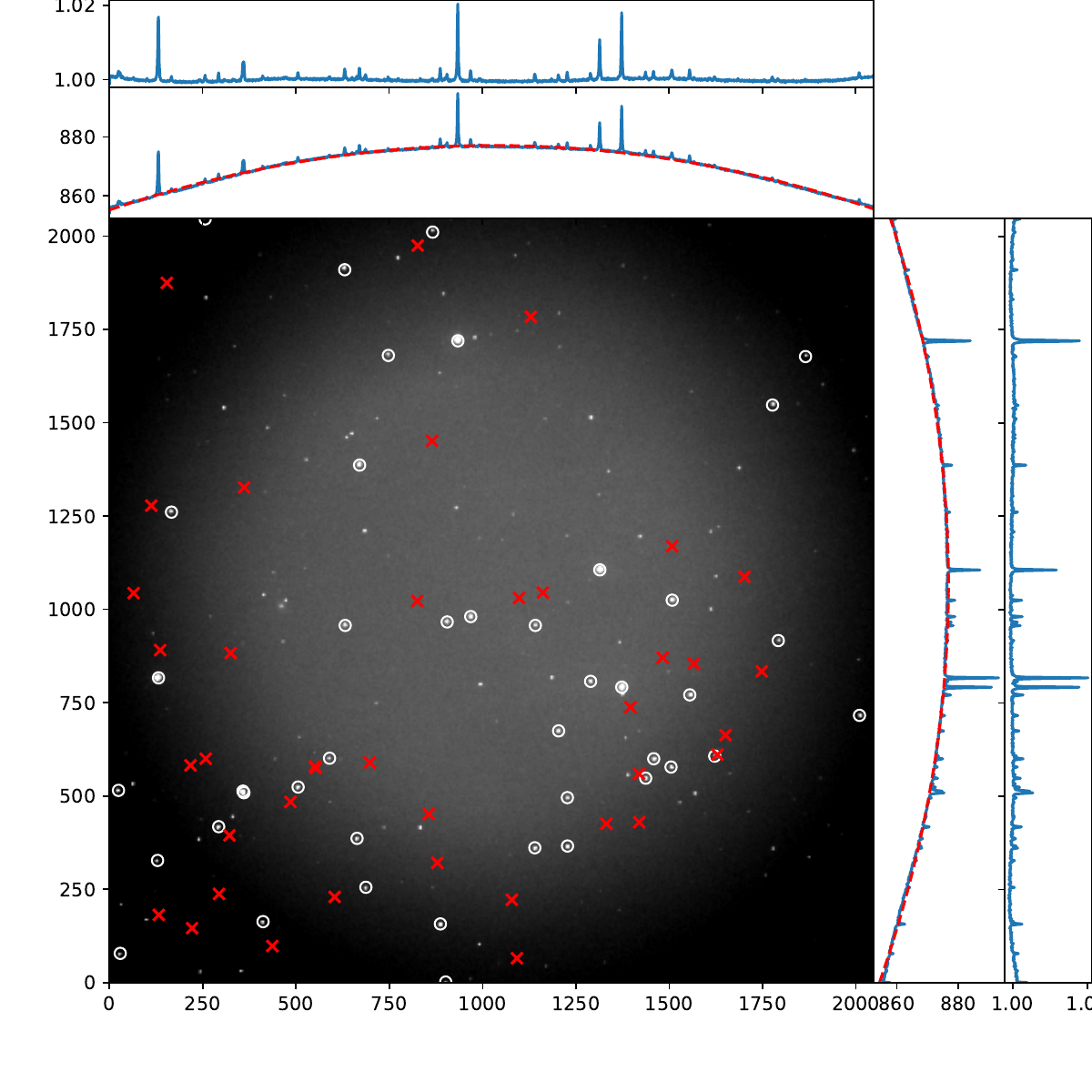}	
	\caption{The reference image, with bright stars (err<0.01) circled out, and positions of the bright stars from the target image also marked with red cross. On the upper and the right, column and row means were plotted, with order-2 polynomial fitted shown in red dashed lines, the normalized curve were shown in the top aadn right.} 
	\label{figImg}%
\end{figure}

In order to show the result of the function, the observation data of a binary named LP UMa were chosen as the sample.
LP UMa was observed with Xinglong 60-cm telescope on Mar 25$^{\mathrm{th}}$ and 26$^{\mathrm{th}}$, 2023, over 1000 photometric images of $V$-, $R$-, and $I$-bands were obtained in each night.
One of images observed in 25$^{\mathrm{th}}$ was randomly chosen as the reference image, and one of images from 26$^{\mathrm{th}}$ as the target image.
Figure \ref{figImg} shows the raw reference image without bias and flat correction, with its column mean shown at the top middle and the row mean at the right middle, and the dotted curves are the fitted background as the normalization factor.
The normalized means were shown at the most top and right panels.

\begin{figure}[h]
	\centering 
	\includegraphics[width=0.5\textwidth]{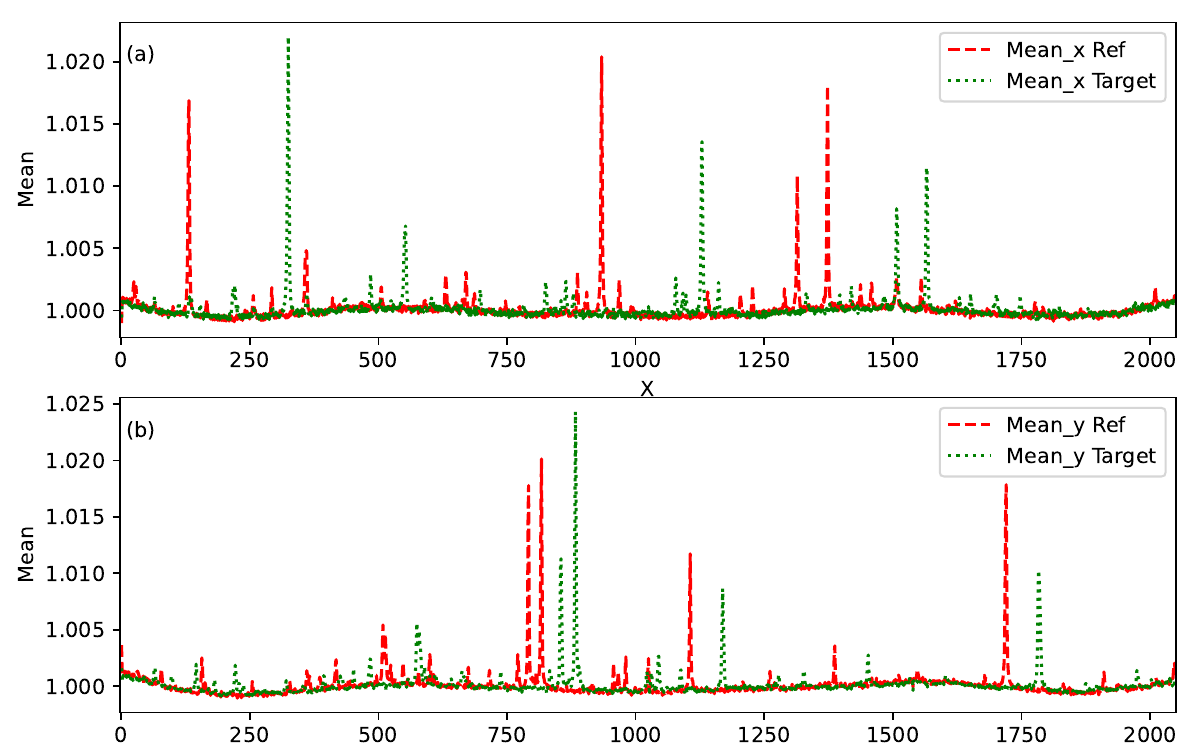}	
	\caption{The column and row means, or mean\_x and mean\_y, of the reference and target images, the curves of the reference image are red dashed curves, while the target as green dotted curves.} 
	\label{figMean}%
\end{figure}

Figure \ref{figMean} shows the normalized means of each column of the reference image and the target image, with the red dashed curve from the reference image, and the green dotted curve from the target image. It can seen that they were similar, but with an offset. 
As comparation, the bright stars extracted from both images are dotted in Figure \ref{figImg}, with stars from the reference image were marked with blue cross, while the red plus indicate the target image. Stars were extracted using the Source-Extractor \citep{1996A&AS..117..393B} and only stars with MAGERR\_AUTO less then 0.01 were selected. Certainly, most stars from both images match, with slightly x and y offsets.

To achieve fast image alignment at the observation site, the above processing uses data that has not been corrected for bias and flat. As a comparison, the corrected data was also processed similarly.

\subsection{Similarity}

The next step is similarity calculation. Calculate the similarity of different offsets by conducting trial translations in different directions.

\begin{figure}[h]
	\centering 
	\includegraphics[width=0.8\textwidth]{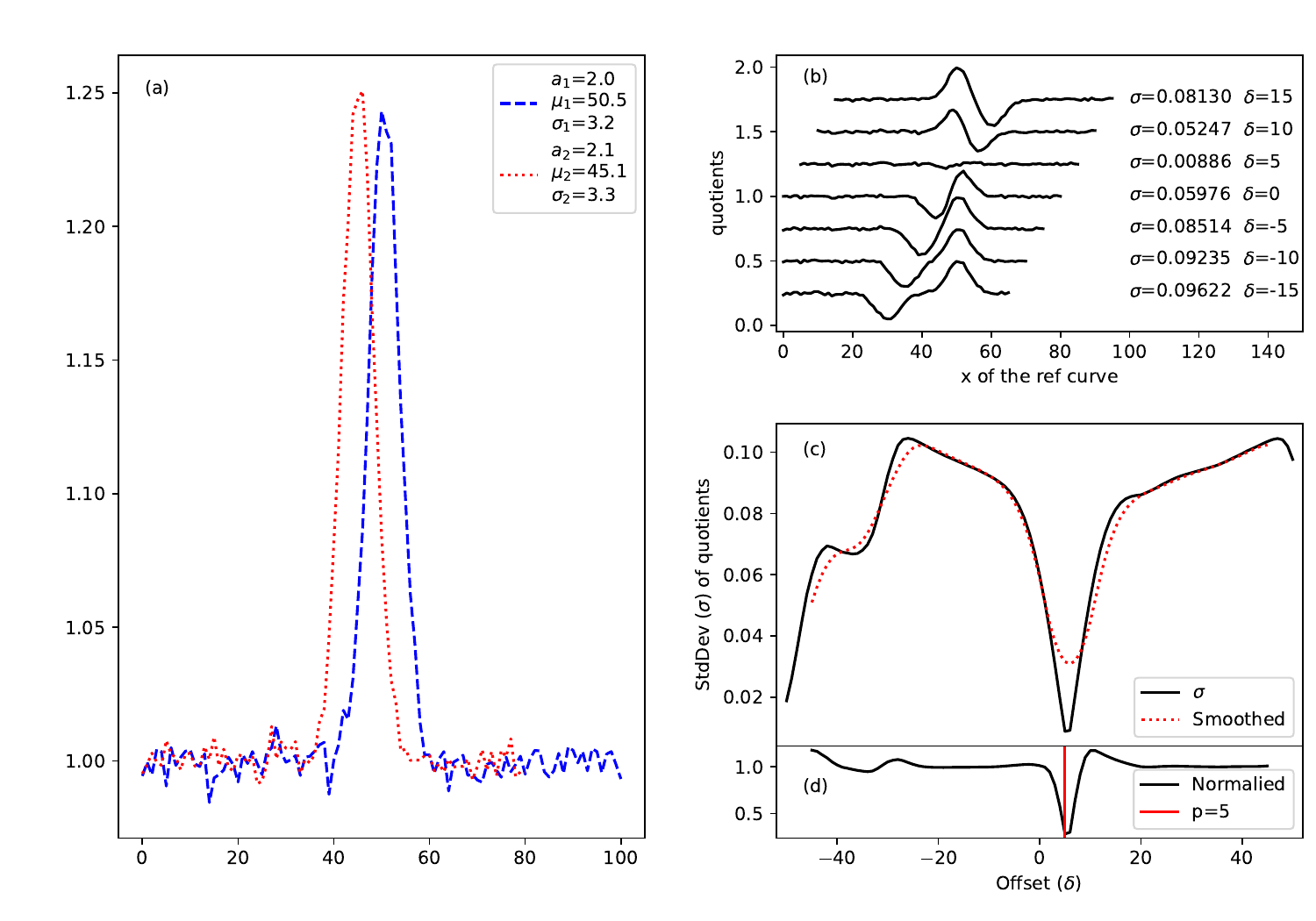}	
	\caption{Simulated Data and Processing, 2 Gaussian curves with noises were shown in panel (a), while the quotients of them was shown in panel (b), with offsets range from -15 to 15 with the step 5. Panel (c) shows the standard deviations of each offsets, and a Gaussian fitting was used to find the minimum position. The valley part of the curve and its fitting were zoomed in a sub-frame.} 
	\label{figSimu}%
\end{figure}

Not like other cross-correlation analysis with sum-product of two arrays (\citep{2008ApJ...677..808Y}), the quotients between the two means were used. If they were the same, the quotients of all points were 1.0, and so the standard deviation of the quotients was 0.0. 
In the real world, the zero standard deviation will only happen when the self-similarity was measured. But the lower standard deviation indicates the more similarity, this is the principle of the algorithm.

To verify the effectiveness of this index, two 1-d Gaussian distribution curves were generated, with slight noise added, as shown in Figure \ref{figSimu}. The $a$, $\mu$, and $\sigma$ of the two curves were (2.0, 50.5, 3.2) and (2.1, 45.1, 3.3), respectively.
Specifically, the two sequences here were intentionally set to different lengths.
The quotients of them was shown in panel (b), with offsets ($\delta$) range from -15 to 15 with the step 5, the standard deviations ($\sigma$) of the quotients were show in the image legends. From the figures, the length of quotients can be seen different, a larger offset shortens the length of the quotient sequence. A too large offset can cause the quotient sequence to be too short, resulting in a low standard deviation, as shown at the both ends of the curve in panel (c). Therefore, in practice, the offset should be within an appropriate range. In fact, if the offset between two images was too large, it is difficult to say that they were observed from the same sky area.
Finally, panel (c) shows the variation of standard deviation with integer offset. 

The minimum value corresponds to an offset of -5, which means that curve 2 should be moved 5 pixels to the left to obtain the best match.

\subsection{Locating the best offset}

Furthermore, from panel (c) of Figure \ref{figSimu}, it can be noted that the curve has a certain global trend, which in practical processing may even lead to locating a fake minimum value. To correct this issue, the curve was normalized and the moving average method was used as the factor. Then the minimum value was located. Panel (d) of Figure \ref{figSimu} shows the normalized curve and the position of the minimum.

Obviously, the results of integers may not be satisfactory. To improve accuracy, the mean sequence can be interpolated to obtain values at 0.1 pixel positions, and then the above process can be performed to obtain higher precious results. Of course, other precious would also work.

\begin{figure}[h]
	\centering 
	\includegraphics[width=0.5\textwidth]{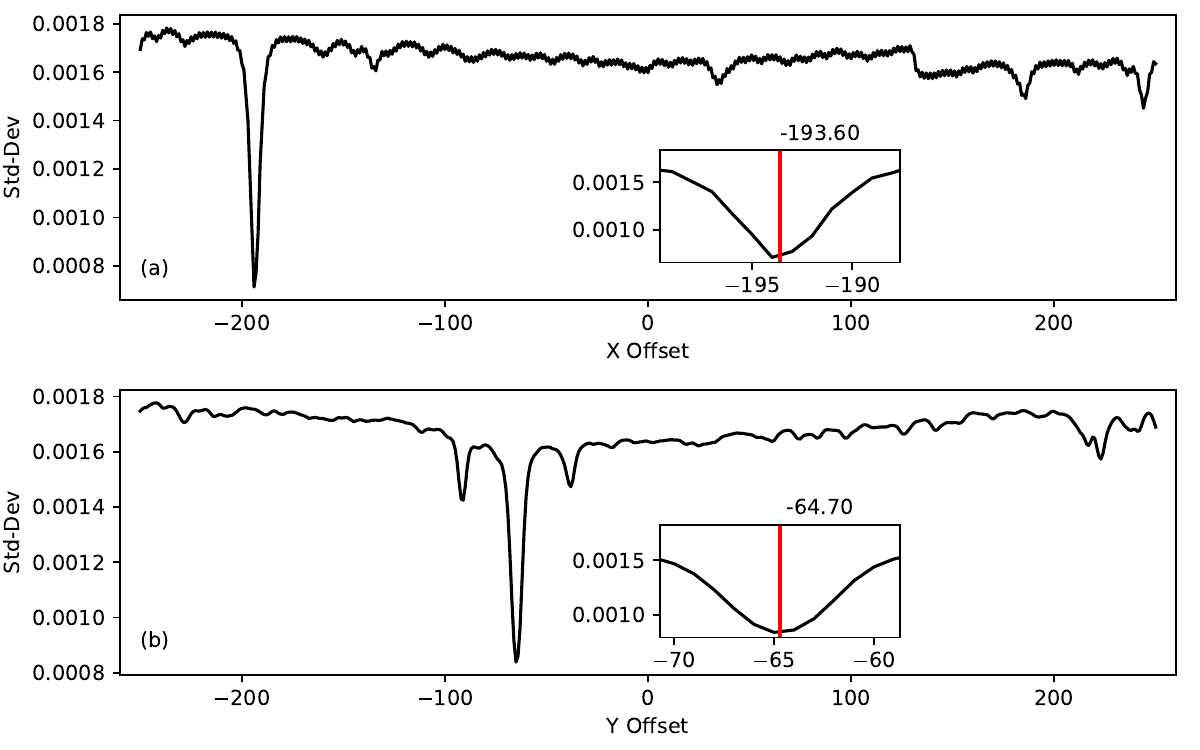}	
	\caption{The standard deviation at different offset, with the valley zoomed in sub-frames.}
	\label{figStd}%
\end{figure}

The above introduces the processing of simulated data, and the following paragraph discusses the effectiveness of this algorithm on real data shown in the previous section.
In simulated data processing, the maximum offset was so large that there was almost no overlap between the data at both ends. When processing actual samples, the maximum offset was selected as about 1/8 of the data width, which was consistent with normal astronomical observations.
The column and row means of the reference and target images is shown in Figure \ref{figMean}. The standard deviation curve calculated using different offsets is shown in Figure \ref{figStd}. Finally, the offsets obtained are -193.5 at x and -64.9 at y.
This can already be used for subsequent works, such as target recognition, and astrometric calibration. For image stacking or subtraction, these values can be used as initial values.

In this algorithm, the x-axis and y-axis are analyzed separately. For equatorial telescopes, the rotation angle of the image is fixed, and there is only translation between the images, so the offsets of the x-axis and y-axis are independent of each other.
For alt-azimuth telescopes, if the image rotation angle is kept consistent during observation, this algorithm can also be used for analysis.

\section{Compare and Application}

In the previous processing, the data did not undergo bias and flat corrections. The results obtained from the corrected data are -195.1 and -65.6 in the x and y directions, respectively, without significant differences.

For astronomical image alignment, it is usually done by finding stars in the image, matching them based on their image coordinates (x, y), and calculating their offset. If necessary, matching may be achieved through triangular relationships of stars.

Next, by comparing the results obtained with traditional methods, the offsets based on the bright stars extracted from the two uncorrected images were -192.89 at x and -64.50 at y. The difference between these two values is acceptable. The offsets based on stars from the corrected images were -194.53 at x and -64.91 at y.
In terms of speed, due to the fact that this algorithm did not require star finding and most of the calculations were simple, this was very fast.

\begin{figure}[h]
	\centering 
	\includegraphics[width=0.5\textwidth]{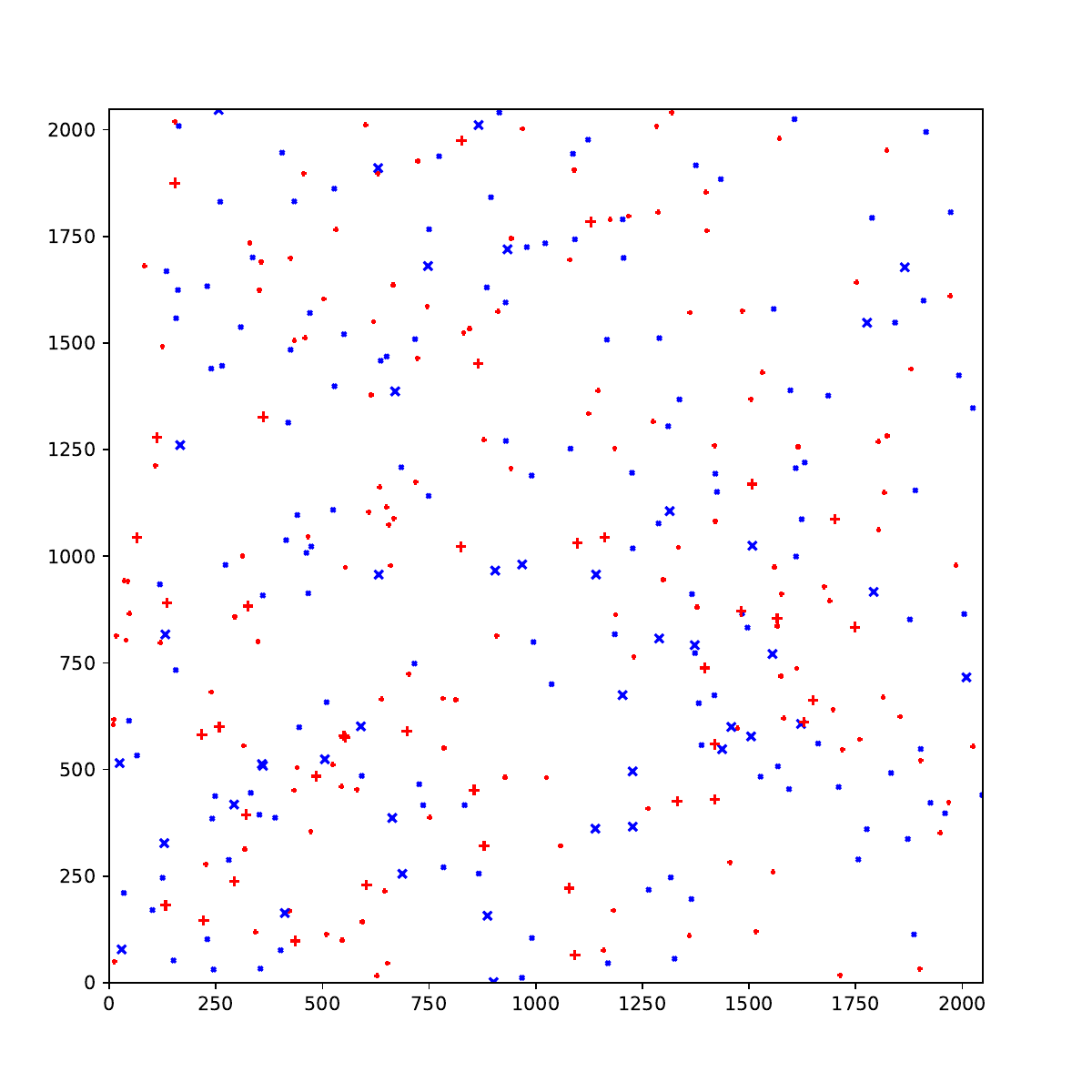}	
	\caption{Stars found from both the reference and target images. Blue crosses came from the reference image while red plus from the target image, and the bright stars  (err $<$ 0.01) were emphasized with larger symbols.}
	\label{figStar}%
\end{figure}

Stars found from both the reference and target images were shown in Figure \ref{figStar}, blue crosses came from the reference image while red plus from the target image, and the bright stars  (err $<$ 0.01) were emphasized with larger symbols. 

\begin{figure}[h]
	\centering 
	\includegraphics[width=0.5\textwidth]{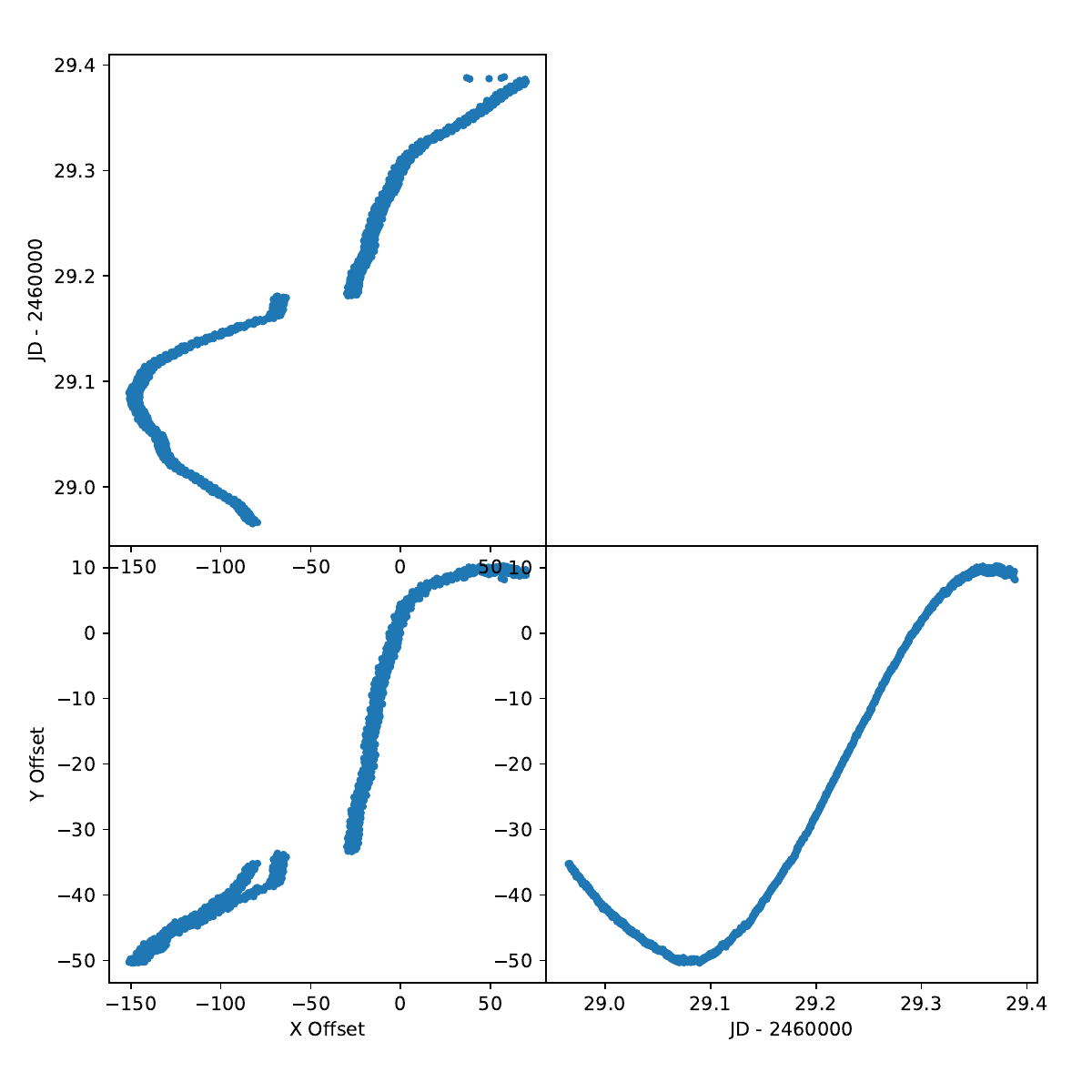}	
	\caption{The offset between images, as well as the offset of the x-axis and y-axis over time.}
	\label{figJump}%
\end{figure}

For the observation data of LP UMa mentioned earlier, this method was used to process over a thousand images observed throughout the night, and all images were perfectly aligned.
Figure \ref{figJump} shows the offset between images, as well as the offset of the x-axis and y-axis over time.

To estimate the speed of this method, and to compare with the astroalign, a popular method, different sizes of images (1K, 2K, and 4K) were processed in jupyter\footnote{https://jupyter.org}, and magic command  ``\%\%time'' was used to show the time cost. The 2K-size images were from Xinglong-60cm  telescope, and the 1K-size images were cutout from the 2K-size images. The 4K-size images were from Nanshan One-meter Wide-field Telescope (NOWT) \citep{}. The commands were shown in the Table \ref{tbCommand} and the results were listed in the Table \ref{tbCompare}.

\begin{table}[]
    \centering
    \begin{tabular}{c l}
        \hline
        Prepare & \texttt{import mean\_offset as mo} \\
         & \texttt{import astroalign as aa} \\
        \hline
        This method & \texttt{\%\%time} \\
         & \texttt{mxa, mya = mo.mean\_xy(imga)} \\
         & \texttt{mxb, myb = mo.mean\_xy(imgb)} \\
         & \texttt{dxab = mo.mean\_offset1d(mxa, mxb)} \\
         & \texttt{dyab = mo.mean\_offset1d(mya, myb)} \\
        \hline
        AstroAlign & \texttt{\%\%time} \\
         & \texttt{transf, (lsta, lstb) = aa.find\_transform(imga, imgb)} \\
        \hline
    \end{tabular}
    \caption{Commands to estimate the time costs}
    \label{tbCommand}
\end{table}

\begin{table}[]
    \centering
    \begin{tabular}{c r r}
        \hline
        Image size & This method & AstroAlign \\
        \hline
        1k & 57.1 ms & 142 ms \\
        2k & 68.8 ms & 361 ms \\
        4k & 116 ms & 2.45 s \\
        \hline
    \end{tabular}
    \caption{Compare the time costs of this methods and the AstroAlign, the time cost listed was the CPU time}
    \label{tbCompare}
\end{table}

\section{Conclusion}

This paper introduces an algorithm for calculating the offset between astronomical photometric images, which is theoretically feasible, fast, reliable, and robust in practical operation. And it has been used for processing actual observation data. However, this algorithm is limited to images that only have translation and is not suitable for comparing images with different rotation angles or different pixel scales.

\section*{Acknowledgements}

This work was supported by the National Key R\&D Program of China No. 2023YFA1608303 and No. 2023YFA1608304. We acknowledge the support of the staff of Xinglong telescopes. This work was also partially supported by the Open Project Program of the CAS Key Laboratory of Optical Astronomy, National Astronomical Observatories, Chinese Academy of Sciences.

\bibliographystyle{elsarticle-harv} 

\end{document}